\documentstyle[aps,psfig]{revtex}

\begin{document}

\draft

\preprint{UFIFT-HEP-01-09,CU-TP-1034}
\title{Cosmology with a long range repulsive force}
\author{Martina Brisudova$^a$\thanks{{\it On leave from Physics Inst., SAS, 
Bratislava.} Electronic address: {\tt brisuda@niobe.iucf.indiana.edu }}, William H.\ 
Kinney$^b$\thanks{Electronic address: {\tt 
kinney@physics.columbia.edu}} and Richard 
Woodard$^c$\thanks{Electronic address: {\tt woodard@phys.ufl.edu}}}
\address{$^a$ Nuclear Theory Center, Indiana University,
2401 Milo B. Sampson Lane, Bloomington, Indiana 47408  \\}
\address{$^b$ Institute for Strings, Cosmology, and Astroparticle Physics,
 Columbia University, New~York,~NY~10027}
\address{$^c$ Dept. of Physics, University of Florida,
 P.O. Box 118440, Gainesville, FL 32611 \\}
\date{\today}
\maketitle

\begin{abstract}
We consider a class of cosmological models in which the universe is filled with
a (non-electric) charge density that repels itself by means of a 
force carried by a vector boson with a tiny mass. When the vector's mass 
depends upon other fields, the repulsive interaction gives rise to an 
electromagnetic barrier which prevents these fields from driving the mass to 
zero. This can modify the cosmology dramatically. We present a very simple 
realization of this idea in which the vector's mass arises from a scalar field.
The electromagnetic barrier prevents this field from rolling down its potential
and thereby leads to accelerated expansion.
\end{abstract}


\section{Introduction}

Recent observations of Type Ia supernovae with high redshift indicate that the
universe is entering a phase of cosmological acceleration \cite{reiss,perlm}.
Identifying the causative agent is perhaps the most exciting task for
fundamental theory at present. There are many candidates. It could be a 
cosmological constant, the need for which
was suggested on the basis of other evidence even before the supernovae
results \cite{KT,OS,LLVW}. Scalars will also work because one can construct 
a potential to support {\it any} homogeneous and isotropic geometry 
for which the Hubble constant does not increase.\footnote{For the 
construction see section 2 of \cite{TW1}.} Minimally coupled scalars becoming
dominant at late times was also suggested before the supernovae results
\cite{PR,Wett,WCL,FJ}. Since then such models have been dubbed 
``quintessence'' \cite{CDS} and have received extensive study
\cite{CLW,ZWS,BM,BCN,Arm}. Nonminimal couplings have also been explored
\cite{APMS} and recent inspiration has been derived from string theory 
\cite{FKPMP,HKS} and extra dimensions \cite{Maeda,HL}. It has even been
suggested that quantum effects may be responsible \cite{LPR}.

In the absence of compelling observational or theoretical support for any of
the existing scenarios it is worth considering what else might be driving 
the late time acceleration we seem to be seeing. We have long advocated that 
there might be interesting cosmological implications from long-range forces 
other than classical gravity \cite{TW,us1}. In the context of a universe which
appears to be pushing itself apart, the repulsive force of vector boson 
exchange naturally comes to mind. That is, suppose the dark matter, or even 
ordinary matter, carries some charge --- not electromagnetic charge although we
will henceforth use that language --- that couples to a $U(1)$ gauge boson 
which we shall refer to as ``the photon.'' With an exactly massless photon the
self-interaction would be infinite, but it can be made finite by the simple 
device of endowing the photon with a tiny mass \cite{paper1}. Since this mass 
can be much smaller than the Hubble constant it is still reasonable to speak of
the resulting force as long-range.

Our purpose here is to survey the cosmology of such theories
to see if they can lead to a late phase of acceleration. 
Therefore we shall not bother with the complications that 
would appear in a fully realistic or completely general model. 
However, it is important to note that there is no shortage of 
plausible candidates for the actual charge carrier, nor any 
lack of physical mechanisms for inducing the vector mass from 
a fundamentally invariant theory. One candidate for our 
charge density is dark matter itself because it is already believed
to exist, although constraints from structure formation make such 
an identification problematic\cite{frieman92}. A less restricted 
alternative, which we adopt here, is a minimally coupled charged 
scalar, similar to the field commonly postulated for ``quintessence'' 
models. The vector might 
acquire its mass, with a fundamentally $U(1)$ invariant Lagrangian, 
through either 
spontaneous or dynamical symmetry breaking. The vector becomes 
massless when some order parameter vanishes. If we also enforce
that increasing the order parameter increases the vacuum energy
then potentially interesting cosmology can arise from the 
tension between the repulsive interaction and the vacuum 
energy. It is neither necessary, nor even desirable
at this stage, to commit to a specific model which illustrates
this tension in a fully realistic fashion.

In this paper we study the cosmological effects of such a long range repulsive 
force. We construct a simple model and analyze it, both analytically and 
numerically, within the framework of a homogeneous and isotropic spacetime.
Our model consists of scalar QED with a nontrivial potential and an explicit 
$U(1)$--breaking term which gives the photon a mass that depends upon the 
scalar's magnitude. Since the global $U(1)$ is preserved, the associated charge is
still conserved, and a homogeneous density of such charge experiences a 
repulsive self-interaction of the type described above. It turns out that the
repulsive self-interaction cannot {\it directly} accelerate the universe, at
least not for very long. However, it can do so {\it indirectly} through 
preventing the scalar from rolling down its potential and thereby driving the
photon mass to zero.

It should be noted that the simple model presented here is identical to
``spintessence'' \cite{GH,BCK} as long as only homogeneous and isotropic field 
configurations are considered. There are potentially important differences when
one allows perturbations because our class of models contains a vector
interaction which spintessence lacks. Note also that the model studied here is
part of a much larger class. It is not necessary to imagine that either the
charge density or the photon mass derive from complex scalars. We consider
this mechanism only because it leads to a simple model in which accelerated
expansion can be shown to occur under certain circumstances.

This paper is organized as follows. In Section II we review how a slightly
 massive vector can be introduced without violating either homogeneity or 
isotropy. We also explain why the mass must depend upon other dynamical 
variables in order to obtain cosmologically interesting late time behavior. In 
Section III we study the simple model described above, first analytically and
then numerically. This leads to the observation, not previously noticed for
spintessence, that the most reasonable initial conditions result in a late time
evolution in which the scalar bounces back and forth between its original
potential and the electromagnetic barrier. We also discuss cosmological parameters.  Section IV contains our summary and conclusions.

\section{Vector long range force}
\label{seclongrangeforce}

A simple model for a repulsive cosmological force is a complex scalar field 
coupled to a $U(1)$ gauge field. For example, one might choose a Lagrangian of 
the form
\begin{equation}
{\cal L}_0 = \sqrt{-g} \left[ g^{\mu \nu} \left(D_\mu \phi\right)^* \left(D_\nu \phi\right)
 - {1 \over 4} g^{\alpha \rho} g^{\beta \sigma} F_{\alpha \beta} 
F_{\rho \sigma} - V\left(\left\vert \phi \right\vert\right)\right]
\; .
\end{equation}
In a homogeneous fluid with a net charge density, the fluid will be 
self-repulsive. However, realizing this situation in an homogeneous, isotropic 
cosmology is problematic. For example, on a closed 3-manifold, the total charge
of any infinite-range force field must be zero, so the charge density must 
vanish. (This is a consequence of the inability to define the ``inside'' and 
``outside'' of an arbitrary Gaussian surface.) On an open manifold, it is 
possible to impose a nonzero charge density, but only at the expense of 
isotropy. One must choose a boundary condition at infinity which selects a 
direction for the lines of force. These obstacles can be evaded by simply 
making the vector massive \cite{paper1}, for example with an explicit Proca 
term:
\begin{equation}
{\cal L} = {\cal L}_0 + {1 \over 2} m^2 g^{\mu \nu} A_\mu A_\nu \sqrt{-g} \; .
\label{eqbrokenlagrangian}
\end{equation}
Note that the current density
\begin{equation}
J_\mu \equiv i e \left[\phi \left(D_\mu \phi\right)^* - \phi^* 
\left(D_\mu \phi\right)\right] \; 
\end{equation}
is still conserved as a consequence of global $U(1)$ invariance.

The unique solution consistent with homogeneity and isotropy has to satisfy 
$\partial_i \phi = 0$ and $A_i = 0$. Homogeneity also implies that $\partial_i 
A_0 = 0$, so that the field strength tensor $F_{\mu \nu} \equiv \partial_\mu 
A_\nu - \partial_\nu A_\mu$ vanishes. The only nontrivial equation of motion 
for the vector field is
\begin{equation}
i e \phi^* \partial_0 \phi - i e \phi \partial_0 \phi^* + A_0 \left(m^2 + 2 e^2 
\phi^* \phi\right) = 0 \; .
\end{equation}
Therefore the unique solution for $A_0$ is
\begin{equation}
A_0 = {i e \left(\phi \partial_0 \phi^* - \phi^* \partial_0 \phi\right) \over 
m^2 + 2 e^2 \phi^* \phi} \; . \label{eqA0solution}
\end{equation} 

The question we wish to ask is: how does a spacetime dominated by such a 
charged scalar evolve? For simplicity, we choose a flat 
Friedmann-Robertson-Walker metric
\begin{equation}
g_{\mu \nu} dx^\mu dx^\nu = dt^2 - a^2\left(t\right) d{\vec x} \cdot d{\vec x}
\; . \label{metric}
\end{equation}
The evolution equation for the metric is the standard Friedmann equation
\begin{equation}
\left(\dot a \over a\right)^2 = {8 \pi \over 3 m_{\rm Pl}^2} T_{00} \; .
\end{equation}
The stress-energy tensor is defined by $T_{\mu \nu} \sqrt{-g} \equiv 2 {\delta 
S}/{\delta g^{\mu \nu}}$ and its nonzero components in this geometry are given by the pressure and energy density,
\begin{eqnarray}
\rho & = & \left(D_0 \phi\right)^* \left(D_0 \phi\right) + {1 \over 2} 
m^2 A_0^2 + V\left(\phi\right) \; , \\
p & = & \left(D_0 \phi\right)^* \left(D_0 \phi\right) + {1 \over 2} m^2 
A_0^2 - V\left(\phi\right) \; .
\end{eqnarray}
We see immediately that the repulsive interaction contributes a term which 
obeys $p = \rho$, rather than the $p < - \rho/3$ needed for acceleration. This
implies that the new term redshifts very rapidly, as $a^{-6}$, and 
quickly becomes negligible. Physically this is because the Universe expands 
while the mass remains constant, so the force eventually becomes short range
on cosmological scales. Therefore, the mere presence of a repulsive force does 
not generically lead to acceleration. It can do so only if the new 
interaction alters the scalar's evolution so as to make the potential $V(\vert
\phi\vert)$ more dominant than it would otherwise have been.

We know that the self-interaction diverges in the massless limit. Imagine a
situation in which the photon mass is not constant but rather depends upon 
some of the other fields. We might expect interesting cosmological effects if 
we begin with a nonzero charge density and then drive the vector mass towards 
zero. In this way the force can remain long range on cosmological scales even 
though the Universe keeps expanding. 

A simple model which allows this behavior can be obtained from the Lagrangian 
(\ref{eqbrokenlagrangian}) by the replacement $m^2 \rightarrow 2\lambda^2 
\phi^2$,
\begin{equation}
{\cal L} = \sqrt{-g} \left[ g^{\mu \nu} \left(D_\mu \phi\right)^* \left(D_\nu \phi\right) 
 +  \lambda^2 \phi^2 g^{\mu \nu} A_\mu A_\nu  - {1 \over 4} 
g^{\alpha \rho} g^{\beta \sigma} F_{\alpha \beta} F_{\rho \sigma} - 
V\left(\left\vert \phi \right\vert\right)\right]  \; . 
\label{eqbarrierlagrangian}
\end{equation}
The homogeneous and isotropic solution (\ref{eqA0solution}) for the vector field becomes:
\begin{equation}
A_0 = {i e \left(\phi \partial_0 \phi^* - \phi^* \partial_0 \phi\right) \over 
2 \left(\lambda^2 +  e^2\right) \phi^* \phi} \; . \label{A0}
\end{equation}
The associated charge density is
\begin{equation}
J^0 = - 2\lambda^2  \phi^* \phi A_0 \; .
\end{equation}
Note that current conservation implies $\partial_0 (a^3 J^0) = 0$.

Homogeneity requires that the scalar depend upon time alone. With this 
simplification its equation of motion becomes
\begin{equation}
\ddot\phi + \left[3 \left({\dot a \over a}\right) - 2 i e A_0\right] \dot\phi + 
{\partial V \over \partial \phi^*} - \left[ \left(\lambda^2 + e^2\right) A_0^2 
+ i e \dot A_0 + 3 i e \left({\dot a \over a}\right) A_0\right] \phi= 0 \; .
\end{equation}
Substituting (\ref{A0}) and making use of current conservation results in the
form
\begin{equation}
\ddot \phi + 3 \left({\dot a \over a}\right) \dot\phi + {\partial V \over 
\partial \phi^*}+ \left(\lambda^2 + e^2\right) A_0^2 \phi = 0 \; .
\label{eqcomplexscalarEOM}
\end{equation}
It is convenient to decompose the scalar into a magnitude and a phase,
\begin{equation}
\phi\left(t\right) \equiv f\left(t\right) e^{i \theta\left(t\right)} \; ,
\end{equation}
so that $A_0$ depends only upon the phase,
\begin{equation}
A_0 = \left({ e \over \lambda^2 +  e^2}\right) \dot \theta \; .
\end{equation}
In these variables the real and imaginary parts of (\ref{eqcomplexscalarEOM}) 
become
\begin{equation}
\ddot f + 3 \left({\dot a \over a}\right) \dot f - \left({\lambda^2 \over 
\lambda^2 + e^2}\right) \dot\theta^2 f + {1 \over 2} V'\left(f\right) = 0 \; , 
\label{eqscalarEOMinter1}
\end{equation}
and
\begin{equation}
\ddot\theta f + \left[2 \dot f + 3 \left({\dot a \over a}\right) f\right]
\dot\theta = {1 \over a^3 f} \partial_0\left(a^3 \dot\theta f^2\right) = 0 \; ,
\end{equation}
Up to coupling constants, the second equation is equivalent to current 
conservation and has the simple solution $\dot\theta \propto  a^{-3} f^{-2}$. 

It is preferable to specify the initial charge density, which is conserved,
rather than the value of the phase. The relation between the two variables is
\begin{eqnarray}
\dot\theta = -  \left(a^3 J^0\right) \left({\lambda^2 + e^2 \over 2 e 
\lambda^2}\right){1 \over a^3 f^2} \; .
\end{eqnarray}
Making this substitution for $\dot{\theta}$ the scalar equation of motion 
becomes
\begin{equation}
\ddot f + 3 \left({\dot a \over a}\right) \dot f - {K \over a^6 f^3} + 
{1 \over 2} V'\left(f\right) = 0 \; , \label{eqEOMf}
\end{equation}
where the constant $K$ is
\begin{equation}
K \equiv \left({\lambda^2 +  e^2 \over 4 e^2 \lambda^2}\right) \left(a^3 
J^0\right)^2 \; .
\end{equation}
We see that the presence of the interaction gives rise to an ``electromagnetic 
barrier'' which prevents the scalar field from relaxing to the origin. 

The energy density and pressure are
\begin{eqnarray}
\rho & = & \dot f^2 + {K \over a^6 f^2} + V\left(f\right) \; , \\
p & = & \dot f^2 + {K \over a^6 f^2} - V\left(f\right) \; . \label{rhop}
\end{eqnarray}
Note that the interaction {\em increases} the energy density, as one would 
expect for a repulsive force. However, the interaction does not, by itself, act
like ``antigravity'' in the popular sense of the phrase. The new term in the 
energy density and pressure looks like a fluid component with equation of state
$p = \rho$, and redshifts like $1 / a^6$. One might therefore expect that the 
interaction becomes negligible at late times. This is indeed true when the 
potential minimum occurs at some nonzero value of $f$. However, the situation 
is more interesting when the minimum is at $f = 0$ because then the scalar is
prevented from rolling down to its minimum by the electromagnetic interaction.
In the next section, we discuss the dynamics of this case.

\section{Dynamics of a model with an ``electromagnetic barrier''}

In this section, we discuss the dynamics of a particularly interesting class of models with an ``electromagnetic barrier'' arising from a vector interaction of the type discussed in Section \ref{seclongrangeforce}. Taking a Lagrangian of the form (\ref{eqbarrierlagrangian}), the equation of motion for the field $f$ is given by Eq. (\ref{eqEOMf}). To complete the description of the 
dynamics, we must also include the Friedmann equation,
\begin{equation}
\left({\dot a \over a}\right)^2 = {8 \pi \over 3 m_{\rm Pl}^2} \left[\dot f^2 + 
{K \over a^6 f^2} + V\left(f\right)\right] \; .
\end{equation}
We are particularly interested in potentials for which the minimum is at $f = 
0$. For this reason, and for simplicity, we will assume a monomial form:
\begin{equation}
V\left(f\right) = V_0 f^b .
\end{equation}
The full set of equations describing the dynamics of the system is
\begin{equation}
\ddot f + 3 \left({\dot a \over a}\right) \dot f - {K \over a^6 f^3} + 
{1 \over 2} b V_0 f^{b - 1} = 0,\label{eqfieldEOM0}
\end{equation}
and
\begin{equation}
\left({\dot a \over a}\right)^2 = {8 \pi \over 3 m_{\rm Pl}^2} \left[\dot 
f^2 + {K \over a^6 f^2} + V_0 f^b\right] \; . \label{eqfullEOM0}
\end{equation}
It is convenient to scale the field $f(t)$ so as to absorb the overall 
$(8 \pi / 3 m_{\rm Pl}^2)$, and then redefine the couplings $K$ and $V_0$ to 
include the residual powers of this factor, 
\begin{eqnarray}
F(t)& \equiv &\sqrt{\left( {8 \pi \over 3 m_{\rm Pl}^2} \right)} f(t) \; , \\
k & \equiv  &\left( {8 \pi \over 3 m_{\rm Pl}^2} \right)^2 K \; , \\
v & \equiv &\left( {8 \pi \over 3 m_{\rm Pl}^2} \right)^{1-{b\over{2}}} V_0\; .
\end{eqnarray}
With these definitions, the equations of motion (\ref{eqfieldEOM0}) and (\ref {eqfullEOM0}) read:
\begin{equation}
\ddot F + 3 \left({\dot a \over a}\right) \dot F - {k \over a^6 F^3} + 
{1 \over 2} b v F^{b - 1} = 0,\label{eqfieldEOM}
\end{equation}
and
\begin{equation}
H^2 \equiv \left({\dot a \over a}\right)^2 = \left[\dot F^2 + {k \over a^6 
F^2} + v F^b\right] \; . \label{eqfullEOM}
\end{equation}

Note that these equations are those of a field moving in an 
effective potential which depends upon the scale factor as well as the scalar
magnitude,
\begin{equation}
V_{\rm eff}\left(F,a\right) \equiv v F^b + {k \over a^6 F^2} \; ,
\end{equation}
(This is similar to the time-dependent potential proposed for Variable Mass Particles or ``VAMPs''\cite{anderson97}.) Due to the dependence upon $a$ the minimum is changing as the Universe 
expands,
\begin{equation}
F_0 =\left[ {2k \over{b v}}\right]^{1 / \left(b + 2\right)}  a^{-6 / \left(b +
2\right)} \; . \label{eqadiabaticminimum}
\end{equation}
Based on this observation, we try to find a self-consistent solution to the equations of motion before proceeding with a numerical analysis.

The strength of the electromagnetic barrier is controlled by the initial 
conditions, i.e. the initial charge density. {\it If} we assume that the 
electromagnetic barrier and the potential energy are each much larger than the 
kinetic terms, then they must approximately balance and  the scalar field has to be approximately $F_0$. Under the same assumption we neglect the kinetic terms in the FRW equation  (\ref{eqfullEOM}), and obtain\footnote{In this paper we use a subscript $0$ (e.g., $H_0$, $a_0$) to indicate the lowest-order solution, not, as is a frequently adopted convention in cosmology, the current value of a quantity.}
\begin{equation}
 H_0^2 \equiv \left[ 
 {k\over{ a^6  F_0^2}} + v 
F_0^b\right]= \left[ 1+ 
 {2 \over{b}}\right] {k \over{a^6  F_0^2}} \  \propto  \ a^{-6 b/ \left(b + 
2\right)}, \label{H0}
\end{equation}
The solution such that $a(t = 0) = 0$ is
\begin{equation}
a_0(t) = a_0(t_0) \left({t\over{t_0}}\right)^{\left(b+ 2\right) / 3 b}.\label{eqmasterscalefactor}
\end{equation}
This solution is indeed self-consistent {\it at late times}. While the potential energy scales like $a^{-6b/(b+2)}$, the kinetic energy drops much faster, like 
$a^{-6}$. While the solution $(F_0,a_0)$ in Eq. (\ref{eqadiabaticminimum}) and Eq. 
(\ref{eqmasterscalefactor}) is {\em a} solution, and is self-consistent, it is 
not  guaranteed that it is {\em the} solution, i.e. a late-time attractor for a range of initial conditions. To answer this question, we resort to numerical solution of the equations of motion.

We have investigated a range of choices for the exponent $b$. 
We find that the solution (\ref{eqmasterscalefactor}) for the time evolution of the scale factor is indeed approached at late 
times for a wide variety of initial conditions. Thus, we are able to model 
practically any kind of cosmological matter which is interacting via a vector 
force by an appropriate choice of~$b$. For example, $b=2$ corresponds to pressureless dust; $b=4$ is radiation. However, of particular interest are scalar potentials that can provide 
an accelerated expansion. Acceleration occurs depending upon the sign of $\rho 
+3p$. Using Eq. (\ref{rhop}) and the scalar's equation of motion (\ref{eqfieldEOM}) we find
\begin{eqnarray}
{8 \pi \over 3 m_{\rm Pl}^2}(\rho + 3 p) = {4\over{a^3}} {d\over{dt}}\left[ a^3
F {d F \over{dt}}\right] - 
2 (1-b) V(F).
\end{eqnarray}
Therefore, acceleration is possible if $b<1$.\footnote{ Of course no potential with $b < 1$ could appear in the Lagrangian of a renormalizable quantum field theory. It might conceivably derive from quantum corrections to the 
effective potential.}
Note that these potentials are very steep as $F \rightarrow 0$, so the electromagnetic barrier is most effective for them. In what follows we will primarily concentrate on this class of potentials. Nevertheless, most of our observations are valid for an arbitrary $b$.

It is important to note, however, that the solution (\ref{eqadiabaticminimum}) for $F_0$ describes the behavior of the field only for very particular boundary conditions. If we assume the solution (\ref{eqadiabaticminimum}) for $F_0$ the Friedmann equation becomes
\begin{eqnarray}
\left({\dot a_0(t_0) \over a_0(t_0)}\right)^2 =&& v \left[F_0\left(t_0\right)\right]^b + {k \over \left[a_0\left(t_0\right)\right]^6 \left[F_0\left(t_0\right)\right]^2}\cr
=&& v \left({b + 2 \over b}\right) \left({2 k \over b v}\right)^{b / \left(b + 2\right)} \left[a_0\left(t_0\right)\right]^{-6 b / \left(b + 2\right)},
\end{eqnarray}
where we have neglected the kinetic term. The initial time is $t_0$. On the other hand, using Eq. (\ref{eqmasterscalefactor}), we have
\begin{equation}
\left({\dot a_0(t_0) \over a_0(t_0)}\right)^2 = \left({b + 2 \over 3 b}\right)^2 t_0^{-2}.
\end{equation}
These two equations imply that the initial value for the scale factor needs to be\footnote{We note that in a flat universe the value of $a$ is arbitrary. We choose this particular scaling because it it convenient for understanding the role of boundary conditions and for numerical evaluation of the equations of motion.}
\begin{equation}
a_0\left(t_0\right) = \left[v t_0^2 \left({b + 2 \over 2 }\right) \left({3 b \over b + 2}\right)^2 \left({2 k \over b v}\right)^{b / \left(b + 2\right)}\right]^{\left(b + 2\right) / 6 b}.\label{eqinitiala}
\end{equation}
Eq. (\ref{eqadiabaticminimum}) then implies that the initial value for the field is
\begin{equation}
F_0\left(t_0\right) = \left[v^{-1} t_0^{-2} \left({2 \over b + 2}\right) \left({b + 2 \over 3 b}\right)^2\right]^{1/b}.\label{eqinitialf}
\end{equation}
Note that the solution (\ref{eqadiabaticminimum}) with this boundary condition is an {\it exact} solution to the equation of motion ({\ref{eqfullEOM}), but only an approximate solution to the Friedmann equation. Numerical analysis indicates that if $t_0$ is large, the solution $(F_0,a_0)$ is a stable solution of the full system of equations. For arbitrary initial conditions, however, the field does not smoothly follow the minimum $F_0$ of the effective potential, but instead oscillates about the minimum. The oscillations do not damp with time. Despite the oscillation, however, the behavior ({\ref{eqmasterscalefactor}) of the scale factor is approximately maintained. 
\begin{figure}
\centerline{\psfig{figure=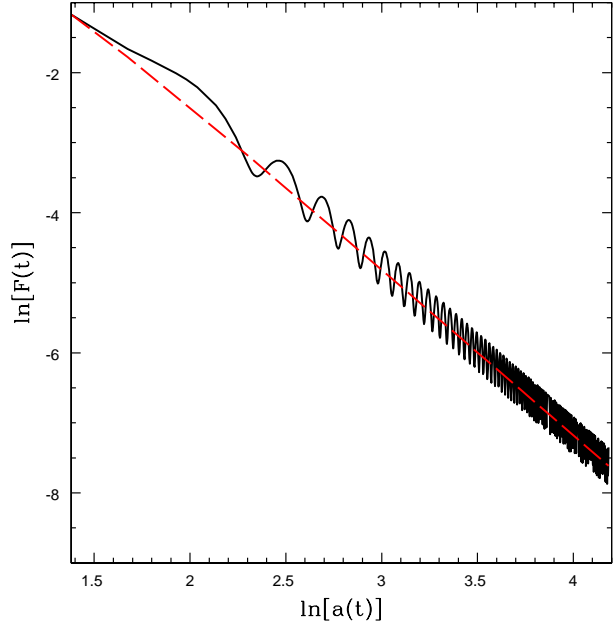,width=3.5in}} \caption{Field $F$ vs. time for $b = 1/2$. The dashed line is the solution (\ref{eqadiabaticminimum}) for $F_0$. }
\end{figure}
Fig. 1 shows the field as a function of 
time for $b=1/2$, $k = 1$, and $v = 0.01$, oscillating about the minimum $F_0$. The boundary conditions are chosen such that the boundary conditions for the field are given by the solution (\ref{eqinitialf}), but the scale factor is set at 0.9 times its value in Eq. (\ref{eqinitiala}). To determine the oscillatory frequency around the minimum $F_0$, consider linearized fluctuations $\Delta F(t)$ and $\Delta a$ around the self-consistent solution $(F_0,a_0)$,  i.e.
\begin{eqnarray}
F(t) & = & F_0(t) + \Delta F(t) \\
a(t) & = & a_0(t) + \Delta a(t).
\end{eqnarray}
The Hubble parameter can then be expanded as
\begin{eqnarray}
H &&= \left({\dot a \over a}\right) \simeq \left({\dot a_0 \over a_0}\right) \left(1 - {\Delta a \over a_0}\right) + {1 \over a_0} {d \Delta a \over d t}\cr
&&\equiv H_0 + H_1,
\end{eqnarray}
where
\begin{equation}
H_1 \equiv {1 \over a_0} \left({d \Delta a \over d t} - H_0 \Delta a\right) = {d \left(\Delta a / a_0\right) \over d t}.
\end{equation}
From (\ref{eqfullEOM}) it follows that the fluctuations  obey the 
following linear equations,
\begin{eqnarray}
 {d^2\over{dt^2}}\Delta F &+& 3 H_0 {d\over{dt}}\Delta F
     + (b+2) {k\over{a_0^6 F_0^4}}  \Delta F
 = - 6 {k\over{a_0^6 F_0^3}}{\Delta a\over{a_0}}\cr
2 H_0 H_1 & = & \dot F_0^2 - 6 {k\over{a_0^6 F_0^2}}{\Delta a\over{a_0}},
 \label{oscil}
\end{eqnarray}
where $H_0(t)$, $a_0(t)$, $F_0(t)$ are the known unperturbed quantities. The $\dot F_0^2$ term is included in the expression because it is the perturbation which prevents the self-consistent solution $(F_0,a_0)$ from solving the Friedmann equation. This term proves in practice to be negligible.

The friction term can be eliminated through the rescaling
\begin{equation}
\Delta F(t) = g(t) a_0^{-3/2}.
\end{equation}
The inhomogeneous equation (\ref{oscil})  for the scalar field $\Delta F(t)$
reduces to
\begin{equation}
d^2g/dt^2 + \omega^2 g = - 6 a_0^{3/2} {k\over{a_0^6 F_0^3}}{\Delta a\over{a_0}},\label{eqoscileom}
\end{equation}
 where the square of the frequency is
\begin{eqnarray}
 \omega^2 = {3\over{4}}\left[- 2 {dH_0\over{dt}}- 3H_0^2\right] + (b+2) {k \over a_0^6 F_0^4}.
\end{eqnarray}
The frequency therefore depends entirely on unperturbed quantities. It is straightforward to show that the first two terms are subdominant for $b < 2$, with
\begin{equation}
H_0^2 \propto \dot H_0 \propto a_0^{-6 b / \left(b + 2\right)},
\end{equation}
so that in the late time limit,
\begin{equation}
\omega^2 = \left(b + 2\right) {k \over a_0^6 F_0^4} = k \left(b + 2\right) \left({2 k \over b v}\right)^{-4 / \left(b + 2\right)} a_0^{6 \left(2 - b\right) / \left(2 + b\right)}.\label{eqlatetimefreq}
\end{equation}
We therefore expect the solution to be of the form of a driven oscillator with frequency given by Eq. (\ref{eqlatetimefreq}). Numerical solution agrees with this analytical result as shown in Fig. 2.

The time dependence of the driving term on the right-hand side of Eq. (\ref{eqoscileom}) can be evaluated as follows. We take the perturbed Friedmann equation (\ref{oscil}) and note that the term
\begin{equation}
\dot F_0^2 \propto a_0^{-6}
\end{equation}
becomes negligible at late times. We then have
\begin{equation}
2 H_0 H_1 = 2 H_0 {d \left(\Delta a / a_0\right) \over d t} \simeq - {6 k \over a_0^6 F_0^2} \left(\Delta a \over a_0\right).
\end{equation}
Since we have the unperturbed solution,
\begin{equation}
H_0^2 = \left({b + 2 \over b}\right) {k \over a_0^6 F_0^2},
\end{equation}
we can write the Friedmann equation as 
\begin{equation}
{d \over d t} \left({\Delta a \over a_0}\right) = - \left({3 b \over b + 2}\right) H_0 \left({\Delta a \over a_0}\right).
\end{equation}
We then have a solution for $\Delta a$,
\begin{equation}
\left({\Delta a \over a_0}\right) \propto a_0^{-3 b / \left(b + 2\right)} \propto t^{-1}.
\end{equation}
The time dependence of the oscillator equation (\ref{eqoscileom}) can then be evaluated,
\begin{eqnarray}
\ddot g + \omega^2 g &&= - a_0^{3/2} {6 k \over a_0^6 F_0^3} \left({\Delta a \over a_0}\right)\cr
&&\propto a_0^{3 \left(6 - 5 b\right) / 2 \left(b + 2\right)}.
\end{eqnarray}
The right-hand side of Eq. (\ref{eqoscileom}) therefore grows with time and drives the oscillations of the scalar field $F$. While we do not have an analytical solution to Eq. (\ref{eqoscileom}), we observe numerically that at late times $\Delta F / F_0 \sim {\rm const.}$, consistent with the rapid growth in amplitude of the oscillations of $g = a^{3/2} \Delta F$. Fig. 2 shows the numerical field solution as a function of the integrated frequency, in agreement with the analytical result. In fact, the driving of the oscillations and the frequency derived through perturbative analysis are quite robust even in the limit where the perturbative expressions above are no longer accurate.
\begin{figure}
\centerline{\psfig{figure=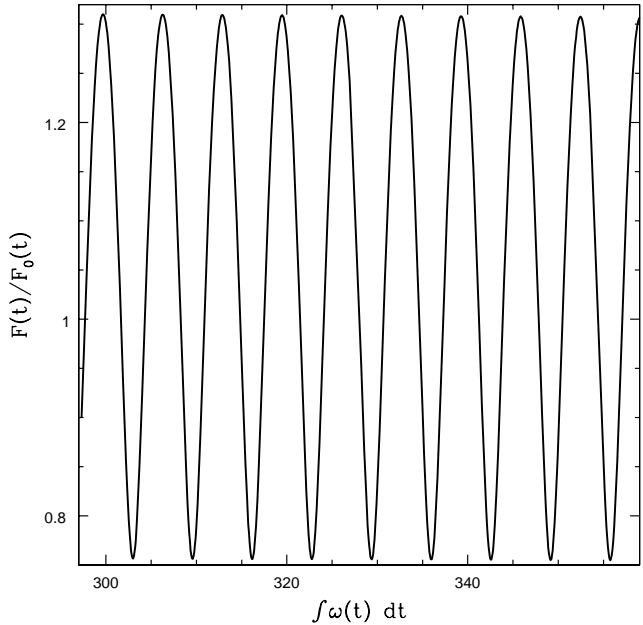,width=3.5in}}
\caption{Field $F(t)$ normalized to the analytic solution $F_0(t)$ (\ref{eqadiabaticminimum}) vs. $\int{\omega(t) dt}$, for $b=1/2$. }
\end{figure}
The most natural initial state is one in which the photon has a large and approximately time independent mass. In this model, this corresponds to the scalar having a large potential energy compared to its kinetic energy and electromagnetic barrier. This particular initial condition leads to an oscillatory late time behavior. The details of the oscillations (i.e. the average value of $F$) depend on the  initial conditions, but the oscillations are generic. The field begins to oscillate even if it starts at the instantaneous minimum of the effective potential, unless the initial condition for the scale factor is given exactly by Eq. (\ref{eqinitiala}). The reason is that the time dependence of the effective potential serves as a driving term for the oscillations. Fig. 3 shows the three components of the stress-energy (potential, kinetic, and barrier) as the field evolves. Although the potential dominates throughout the evolution, all three terms scale identically with $a$. The precise relationship between the three components is very sensitive to initial conditions. Note in particular the fact that the kinetic energy scales identically with the other terms in the stress-energy, unlike in the case of the self-consistent solution $F_0$.

\begin{figure}
\centerline{\psfig{figure=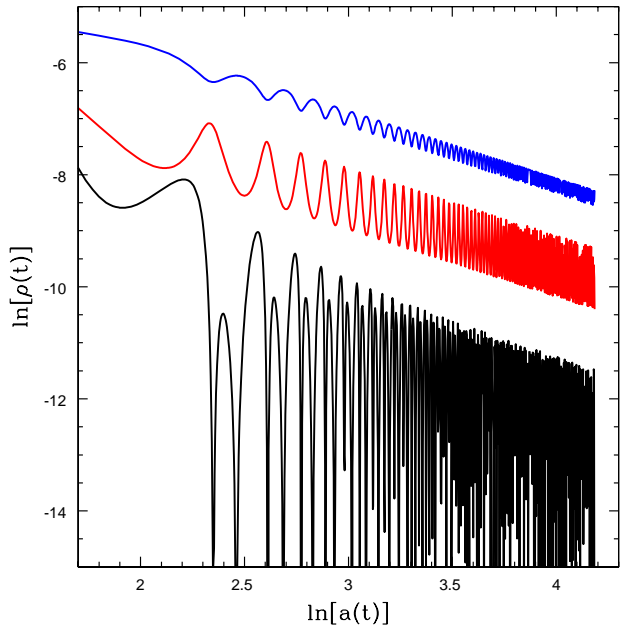,width=3.5in}}
 \caption{The three components of the energy density vs. scale factor for $b = 
 1/2$. The potential (upper line, blue) and ''barrier'' (middle line, red) terms dominate the energy density, while the kinetic term (lower line, black) is subdominant. All three scale identically with $a$.}
 \end{figure}

Of most interest are, of course, the cosmological parameters. Do the small 
oscillations of the field affect the background? The deceleration parameter $q$ follows the behavior of the field (Fig. 4). At times when the field is at its maximum, $q$ reaches its minimum at $-0.7$. Note in particular that despite the fact that the kinetic energy of the field vanishes when $q$ is at its minimum, the equation of state never reaches $p = -\rho$ due to the contribution of the barrier term to the stress-energy, and $q > -1$ at all times. The system, however is on average in a state of accelerated expansion.

\begin{figure}
\centerline{\psfig{figure=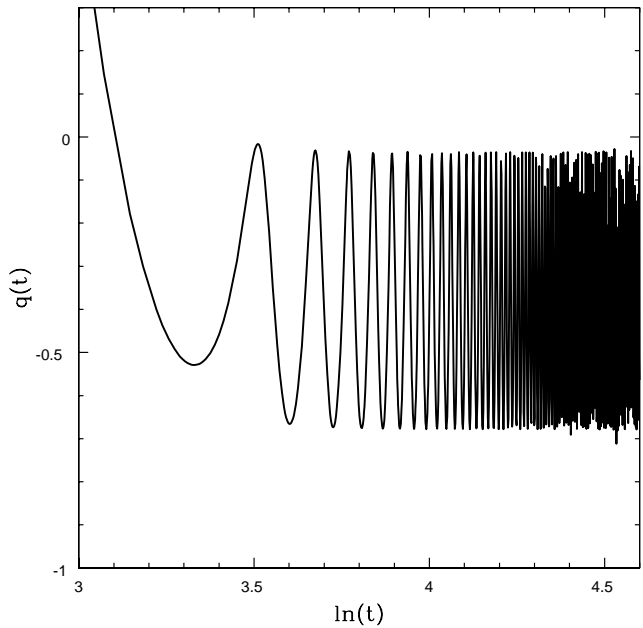,width=3.5in}}
\caption{The deceleration parameter $q$ vs. time. The time variable is in the internal units of the numerical solution, $t = [20, 100]$}
\end{figure}

The behavior of the Hubble constant for the same set of initial conditions is shown in Figure 5. It, too, exhibits oscillations. Although the specifics of the field evolution are model dependent, they arise from the field oscillating about the minimum of the effective potential created by the potential and barrier terms. So, oscillations can be expected in any model of this type.  Therefore the anomalous jitters in the Hubble constant and deceleration parameter can serve as a distinguishing feature of such models. A possible observational signature of such behavior would be the presence of scatter on a Hubble diagram which cannot be attributed to statistical error.

\begin{figure}
\centerline{\psfig{figure=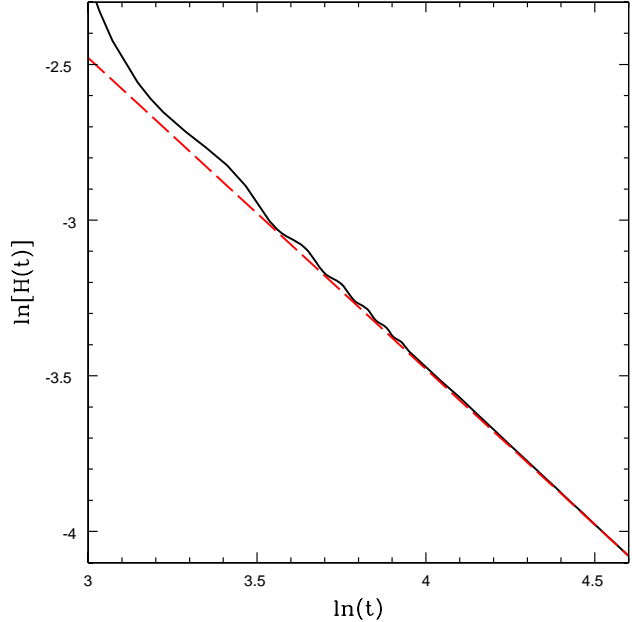,width=3.5in}}
\caption{The Hubble parameter vs. time. The (red) dotted line
is the result for the self-consistent solution.}
\end{figure}

The Hubble parameter itself is not directly observable, however. Cosmological expansion is measured by comparing the redshifts and distances of standard candles such as Type Ia supernovae. The luminosity distance measured by the observed brightness of a known standard candle, is
\begin{equation}
d_{\rm L} = a\left(t_0\right) (1 + z) \int_{0}^{z}{d z' \over H\left(z'\right)},
\end{equation}
where $t_0$ is the current time, and z is the redshift of the source. This is most commonly expressed as the distance modulus, or dimming (in magnitudes) of the observed object:
\begin{equation}
\left(m - M\right) = 5 \log\left(d_{\rm L}/10\,{\rm pc}\right) = 5 \log\left(H_0 d_{\rm L}\right) + 41.6.
\end{equation}
Here $M$ is the absolute magnitude of the standard candle, $m$ is the observed magnitude, and the luminosity distance is expressed in units of the Hubble constant today, $H(t_0) d_{\rm L}$. We have taken $H(t_0) = 70\,{\rm km/sec/Mpc}$. Since we consider a toy model in which only the oscillating scalar contributes to the energy density of the universe, the choice of what to label the current time $t_0$ is arbitrary. (A more realistic model would consist of both matter and scalar field, and the current time would be fixed by the observed ratios of their densities, $\Omega_{\rm M} \sim 0.3$, $\Omega_{\phi} \sim 0.7$.) We take an optimistic choice for $t_0$, such that the early, large oscillations in the Hubble parameter are at a redshift of approximately 1 (which in the units used in the numerical simulation, is $t_0 = 68$.) Fig. 6 shows a plot of distance modulus vs. redshift for the Hubble parameter plotted in Fig. 5. Despite the strong oscillations in the deceleration parameter, the oscillations on the Hubble diagram are very small, in fact too small to be evident on the plot.
\begin{figure}
\centerline{\psfig{figure=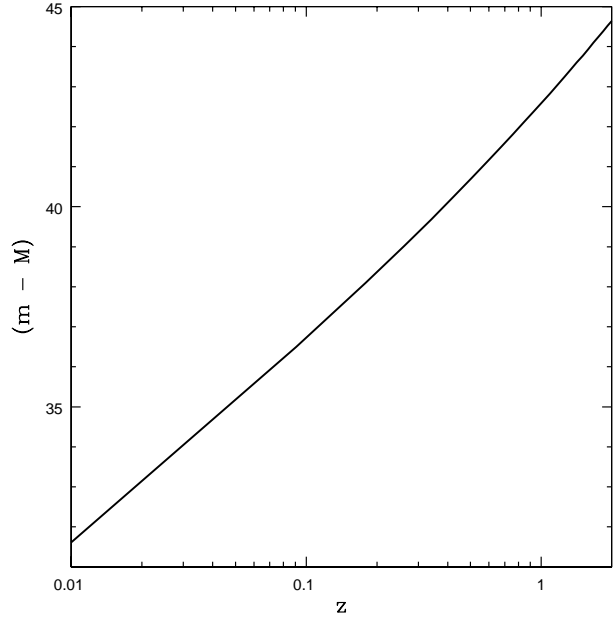,width=3.5in}}
\caption{Distance modulus vs. redshift. Note that the oscillations in the Hubble parameter are not visible in this graph.}
\end{figure}
The magnitude of the oscillations on the Hubble diagram can be shown by plotting the residual $\Delta(m - M)$ between the exact solution and the self-consistent solution, shown in Fig. 7. Fig. 8 shows a plot of the deceleration parameter $q$ as a function of redshift for the same choice of $t_0$. 
\begin{figure}
\centerline{\psfig{figure=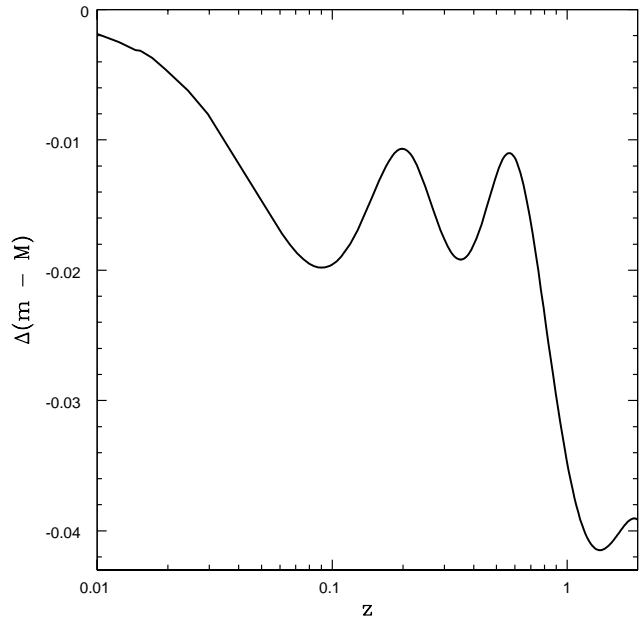,width=3.5in}}
\caption{Residual $\Delta(m - M)$ vs. redshift, relative to the self-consistent solution. The oscillations are visible at the level of $\Delta(m - M) \sim 0.04$.}
\end{figure}
\begin{figure}
\centerline{\psfig{figure=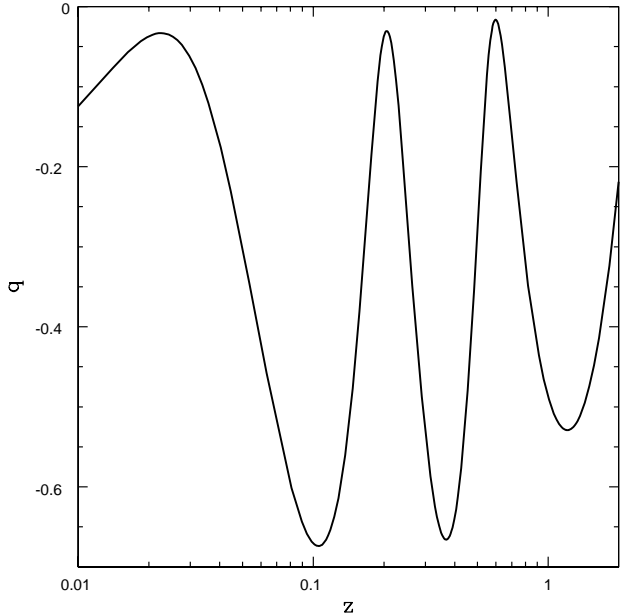,width=3.5in}}
\caption{Deceleration parameter $q$ vs. redshift, for reference to Figs. 6 and 7.}
\end{figure}
We see that the magnitude of the oscillations is $\Delta(m - M) \sim 0.04$,  smaller than the accuracy of current supernova Ia observations, which have errors on the order of $\Delta(m - M) \sim 0.15\ -\ 0.20$\cite{SNIa}. Other choices of initial conditions result in a similar order of magnitude for $\Delta(m - M)$, to within a factor of a few. Detection of such a signature would present a formidable observational challenge, but it is conceivable that future measurements such as those by SNAP\cite{SNAP} could approach this level of accuracy. We note with interest that there is in fact some evidence of small scatter beyond observational uncertainties in existing data, at the level of .12 magnitude\cite{SNIa}.

\section{Summary and Conclusion}

In this paper we studied how a presence of a long range repulsive force affects
cosmology. We have explicitly constructed a model for a cosmology dominated by a charged scalar field with a long-range repulsive interaction. The simplest such model, one with an explicitly broken $U(1)$ gauge interaction (\ref{eqbrokenlagrangian}) is sufficient to show that a universal repulsive interaction does not generically introduce a negative pressure, but rather a term with equation of state $p = \rho$. Thus, a universal repulsive force by itself does not act like a cosmological constant. It can have an interesting effect on cosmology, including support of inflation, but in a more subtle way.

In order for the repulsive force to be of any significance at late times,
its range has to be allowed to increase, i.e. the mass of the force carrier 
needs to be dynamical. Then because of charge conservation, an electromagnetic 
barrier arises \cite{paper1} (in addition to a kinetic barrier as in 
spintessence \cite{GH,BCK}). The system behaves as a driven oscillator around the effective potential arising from the sum of its original potential and the electromagnetic barrier. Even though the resulting electromagnetic barrier contributes a $w=+1$ term to the stress tensor it can lower the deceleration parameter $q$ by pushing the field back up its potential. For concave potentials, the deceleration parameter can even be negative. 

Here we have chosen 
the {\it simplest} case of a dynamically changing mass, that is, one without any additional fields required. Because of this choice, the electromagnetic barrier has the same form as the kinetic energy barrier and they can be combined. The resultant equations of motion thus {\it look} like spintessence. An important distinction arises when spatial perturbations are considered. In our model this would generate a nonzero electric and magnetic fields which would tend to resist clumping of the charges. Therefore, it might distinguish our model from these models with respect to their tendency to decay into Q-balls \cite{Kasuya}.

We study the model both analytically and numerically. Our analytical formulas are confirmed by the numerical simulations. We find that there exists
a self-consistent solution such that kinetic energy is negligible compared to 
the potential energy of the scalar field and its electromagnetic barrier 
(identical, in this particular model, to the spintessence solution). 
However, for generic initial conditions, the solution displays oscillatory behavior. We find that the small oscillations change scaling of the kinetic energy, so that it scales in the same way as the potential and barrier terms in the stress-energy. Depending on initial conditions, the kinetic energy can be of the same order of magnitude as the other terms in the stress energy, unlike the self-consistent solution. 

The presence of oscillations can result in interesting observable consequences. In particular, the Hubble parameter and the deceleration parameter will also in general oscillate, creating an anomalous ``spread'' in a Hubble diagram not explainable by statistical error. Depending upon
how big the effect is, it may or may not be observable and it may possibly 
constitute a signature of a long range repulsive force. 

In conclusion, we have presented the simplest of a class of models with a long 
range vector force.
We emphasize that  the key feature that makes our model work 
is the electromagnetic potential becoming  increasingly long range. This is 
achieved by driving the mass of the vector dynamically to zero. However, apart 
from the simple case presented here, the same effect can be achieved through a 
wide range of possibilities.  For 
example, we can have another scalar field generate the
photon mass. We can have it generated by dynamical symmetry breaking
of a Fermi field. We can even have the charge density provided by a
Fermi field condensate. 
Nevertheless, many of the features present in this specific model, including the anomalous jitters, can be expected to be generic for  models of this type.

\section*{Acknowledgments}
This work was partially supported by DOE contract DE-FG02-97ER-41029 and by
the Institute for Fundamental Theory. WHK is supported by the Columbia 
University Academic Quality Fund. ISCAP gratefully acknowledges the generous
support of the Ohrstrom Foundation.

\end{document}